\documentclass[aps,twocolumn,showpacs,preprintnumbers,amsmath,amssymb,superscriptaddress]{revtex4}
\usepackage[utf8x]{inputenc}
\usepackage{ucs}
\usepackage{amsmath}
\usepackage{amsfonts}
\usepackage{amssymb}
\usepackage{graphicx}
\usepackage{setspace}

\begin{document}

\title{Strain-induced topological insulator phase transition in HgSe}
\author{Lars Winterfeld}
\affiliation{Department of Physics, California State University, Northridge, California 91330-8268, USA}
\affiliation{Institut für Physik, University of Technology Ilmenau, 98684 Ilmenau, Germany}
\author{Luis A. Agapito}
\affiliation{Department of Physics, California State University, Northridge, California 91330-8268, USA}
\author{Jin Li}
\affiliation{Department of Physics, California State University, Northridge, California 91330-8268, USA}
\author{Nicholas Kioussis}
\email[E-mail at:~] {nick.kioussis@csun.edu}
\affiliation{Department of Physics, California State University, Northridge, California 91330-8268, USA}
\author{Peter Blaha}
\affiliation{Institute for Materials Chemistry, TU Vienna, A-1060 Vienna, Austria}
\author{Yong P. Chen}
\affiliation{Department of Physics, Purdue University, West Lafayette,  IN 47907, USA}

\date{\today}

\begin{abstract}
Using {\it ab initio} electronic structure calculations we investigate the change of the band structure and the $\nu_0$ topological invariant in HgSe
(non-centrosymmetric system)  under
two different type of uniaxial strain along the [001] and [110] directions, respectively.
Both compressive [001] and [110] strain leads to the opening of a (crystal field) band gap (with a maximum value of about 37 meV) in the vicinity of $\Gamma$, and
the concomitant formation  of a camel-back- (inverse camel-back-) shape valence (conduction)
band along the direction perpendicular to the strain with a minimum (maximum) at $\Gamma$. We find that the $\mathbb{Z}_2$ invariant $\nu_0$=1
which demonstrates conclusively that HgSe is a strong topological insulator (TI).
With further increase of the strain the band gap decreases vanishing at a critical strain value (which depends on the strain type) where HgSe undergoes a transition from a strong TI to
a trivial (normal) insulator.  HgSe exhibits a similar behavior under a tensile [110] uniaxial strain. On the other hand, HgSe remains a normal insulator
by applying a [001] tensile uniaxial strain.
Complementary electronic structure calculations of the non-polar (110) surface under compressive [110] tensile strain show two
Dirac cones at the $\Gamma$ point whose spin chiral states are associated with the top and bottom slab surfaces.
 \end{abstract}

\maketitle

\section{Introduction}
There has been an explosion of theoretical and experimental research in the search for three-dimensional (3D) topological insulators (TIs),
which are a new quantum state of matter with promising applications
in spintronics and multifunctional topological-based devices and interfaces \cite{HasanKane,Fu&Kane07}.
TIs are characterized by gapped bulk states, similar to a normal (trivial) insulators,
with robust gapless {\it topologically protected} metallic surface states resulting from the strong spin-orbit coupling (SOC).
This in turn leads to an odd number of Dirac-cone-like dispersions crossing the Fermi level at the surface and to a non-trivial $\mathbb{Z}_2$ topological invariant.
The surface states exhibit an unconventional spin texture with the spin direction locked to the two-dimensional (2D) electron momentum.
Current ongoing research efforts focus primarily on  the prototypical single Dirac-cone family of Bi$_2$Te$_3$, Bi$_2$Se$_3$, and Sb$_2$Te$_3$ 3D TIs,
where angle-resolved- photoemission experiments have shown the chiral spin texture of the surface states\cite{Zhang,Xia,Chen}.

Although the band inversion near the time-reversal invariant momenta (TRIM) points
($\vec{k}\equiv-\vec{k}$ mod $\vec{G}$, where $\vec{G}$  is a reciprocal lattice vector), is a strong indication that a system is TI,
it is not definitive because the topological invariant is a global property of the entire occupied Brillouin zone (BZ).
Thus, in order to confirm conclusively whether a 3D system is topologically trivial or non-trivial it is essential to compute the $\mathbb{Z}_2$ invariant.
For band insulators with crystal inversion symmetry,
the topological invariant number can be easily computed as the product of half of the parity
(Kramers pairs have identical parities) numbers for all the occupied states at the TRIM points\cite{Fu&Kane06}.
For the general case of non-centrosymmetric systems (such as zinc-blende structures)
the topological invariant can be determined either by integrating both the Berry's connection and curvature over half of the BZ\cite{Fukui}  or
by calculating the evolution of the Wannier function center during  a ``time-reversal pumping" process\cite{Yui}.
These two equivalent approaches are discussed in detail in Sec. II.

The mercury chalcogenides HgSe and HgTe in the zinc-blende (ZB) structure belong to a group of unique materials exhibiting the so-called inverted band structure.
In this band structure, the energy of the anion $s$-derived $\Gamma_6$ state,
which is the conduction-band minimum (CBM) at $\Gamma$ for most ZB semiconductors,
is below the anion $p$-like valence-band maximum (VBM) with $\Gamma_8$ symmetry.
However, the experimental results for the electronic structure of HgSe are controversial.
Early photoelectron spectroscopic experiments\cite{Gawlik,Truchsess} on n-type HgSe indicated that
this material should have a gap with a positive value for $E_0 \equiv E(\Gamma_6) - E(\Gamma_8)$  = 0.42 eV.
On the other hand, photoemission experiments\cite{Janowitz} of the non-polar HgSe (110) surface (to minimize the effects of surface charges) could not unequivocally determine
the sign of $E_0$ due to lack of resolution of the sequence of the $\Gamma_6$, $\Gamma_7$ and $\Gamma_8$ bands.

Currently, the consensus of a large amount of experimental results\cite{Einfeldt} for HgSe seems to give the sequence of the energy bands
in ascending order is $\Gamma_6$, $\Gamma_7$ and $\Gamma_8$, with $E_0 \approx$ -0.2 eV and
the spin-orbit splitting, $\Delta_{SO} \equiv E(\Gamma_8) - E(\Gamma_7) \approx$ 0.4 eV.
{\it Ab initio} calculations based on the GW (G is the Green's function and W is the screened Coulomb interaction)
scheme find\cite{Svane,Sakuma,Fleszar,Delin} $E_0 \approx$ -0.32 to -0.58 eV and $\Delta_{SO}\approx$ 0.27 to 0.32 eV, in agreement with experiment.

In contrast, in HgTe the sequence of the energy bands is $\Gamma_7$, $\Gamma_6$, and $\Gamma_8$\cite{Svane}.
Interestingly, recent transport experiments\cite{Brune} on a 70-nm-thick HgTe film, strained by epitaxial growth on a CdTe substrate,
show that the strain induces a band gap in the otherwise semi metallic HgTe, rendering it a 3D TI. Similar strain appears when
thin HgSe films are epitaxially grown on CdTe or CdSe substrates.


The objective of this work is to investigate the effects of uniaxial strain on the topological phase of HgSe employing {\it ab initio} calculations.
The rest of this paper is organized as follows:
 Section II describes the methodology used to calculate topological invariants and the band structures.
In section III we analyze the evolution of the band structure and the topological invariant number under two different type of uniaxial
strain along the [001] and [110] directions, respectively. In order to corroborate the topological insulator nature we also present
results for the electronic structure of the non-polar (110) HgSe surface and the spin polarization of the surface states.
Finally, conclusions are summarized in Sec. IV.

\section{Methodology}
For 2D systems, there are two commonly employed numerical approaches to determine whether a non-centrosymmetric insulator is a TI.
\begin{enumerate}
\item {
Use the periodic part of the Bloch functions, $|u_n(\vec{k})\rangle$,
to calculate the Berry connection, $\vec{A}(\vec{k}) = i \sum_n\langle u_n(\vec{k})|\nabla_{\vec{k}}|u_n(\vec{k})\rangle$, involving the sum over the occupied bands.\cite{Fukui}
For a 2D system, both the Berry connection and the Berry curvature, $\mathfrak{F}(\vec{k}) = \nabla_{\vec{k}}\times\vec{A}(\vec{k})$,
can then be used to directly compute the topological invariant, $z$, involving the integration over half of the BZ, denoted by ${\mathfrak{B}}^+$, i.e.,
\begin{equation*}
z = \frac{1}{2\pi} \left[\oint_{{\partial}{\mathfrak{B}}^+} \vec{A}(\vec{k})\cdot d\vec{k} - \int_{\mathfrak{B}^+}\mathfrak{F}(\vec{k}) d^2k\right] \text{mod 2}.
\end{equation*}
}

\item {
 Employ a recently proposed equivalent method for the $\mathbb{Z}_2$ topological invariant based on the U(2N) non-Abelian Berry connection\cite{Yui}.
 This approach allows the identification of the topological nature of a general band insulator
 without any of the gauge-fixing problems that plague the concrete, previous implementation of invariants.
 The main idea of the method is to calculate the evolution of the Wannier function center (WFC) directly during a ``time-reversal pumping" process,
 which is a $\mathbb{Z}_2$ analog to the charge polarization proposed by Fu and Kane\cite{Fu&Kane06}.
 The evolution of the WFC along $k_y$ corresponds to the phase factor, $\theta$,
 of the eigenvalues of the position operator, $\mathbb{\hat{X}}$, projected into the occupied subspace.
 Each state of the $n$th occupied band is indexed by three quantum numbers $| n,\, k_x,\, k_y \rangle$.
This allows to define a square matrix $F(k_x, k_y)$ containing the overlap integrals,
\begin{equation*}
\left( F(k_x, k_y) \right)_{mn} = \langle m, k_x, k_y | n, k_x + \Delta k_x, k_y \rangle \;,
\end{equation*}
where $\Delta k_x = \frac{2\pi}{N_x a}$ is the discrete spacing of $N_x$ points.
One can in turn calculate the complex unitary square matrix

\hspace{1.8cm}$D(k_y) = \prod\limits_{j=0}^{N_x-1} F\left(j \, \Delta k_x, k_y\right) \;$,

which has the complex eigenvalues $\lambda_l(k_y) = |\lambda_l(k_y)| e^{i \theta_l(k_y)}$.
 The topological invariant $z$ is then calculated by the even or odd number of crossings of any arbitrary horizontal ($\theta$=const.) reference line with the evolution of the $\theta_l$'s, mod 2.
 }
\end{enumerate}
For a 3D bulk system, it is necessary to calculate two different invariants, $z_{0}$ and $z_{\pi}$, for the two different BZ planes, $k_z$ = 0 and $k_z$ = $\pi$, containing
eight TRIM points. The computation for each plane is analogous to the 2D case which has four TRIM points.
We employ the second aforementioned approach
to calculate the 3D topological invariant, $\nu_0$, which is defined as $\nu_0 \equiv (z_{0} - z_{\pi}$) mod 2.
The system is in the strong topological insulator (trivial insulator) phase, if $\nu_0$ is 1 (0).
More specifically, the system is a non-trivial or strong TI if the evolution curves of the WFC cross an arbitrary reference line an odd number of times
in the $k_z$ = 0 plane ($z_0$ = 1) and an even number of times in the $k_z$ = $\pi$ plane ($z_{\pi}$ = 0), or vice versa\cite{Yui}.
On the other hand, if the evolution curves cross the arbitrary reference line an even ($z_{0}$=$z_{\pi}$=0)
or odd ($z_{0}$=$z_{\pi}$=1) number of times in {\it both} the $k_z$ = 0 and $k_z$ = $\pi$ planes, the system is topologically a trivial or normal insulator (NI).


The {\it ab initio} calculations employed the Vienna Ab initio Simulation Package (VASP) code\cite{Kresse,Kresse2} employing
the Perdew-Burke-Ernzerhof (PBE) flavor of the exchange-correlation functional\cite{Perdew} and the projected-augmented-wave approach\cite{Blochl} to represent the electron-ion interaction.
The spin-orbit coupling was included in the self-consistent calculations.
The Hg 5$d$ electrons are treated explicitly as valence electrons.
The energy cutoff of the plane-wave expansion of the basis functions was set to be 550 (450) eV for the bulk (surface) calculations.
The BZ integration was performed with a 11$\times$11$\times$11 (7$\times$7$\times$1) \textbf{k}-point Monkhorst-Pack mesh\cite{Monkhorst} for the bulk (surface) calculations.
The equilibrium volume, $V_0 = 61.86 \text\AA^3$ per formula unit, is in good agreement with previous {\it ab initio} calculations\cite{Radescu}.
The experimental lattice constant a=6.084 \AA\,\cite{Radescu} differs by 3\% from the theoretical equilibrium of PBE.
The non-polar (110) surface is modeled by a periodic slab consisting of 21 atomic layers with a 12 {\AA}-thick vacuum region separating the periodic slabs.

An accurate description of the electronic structure is a prerequisite in the search and discovery efforts for next-generation TIs.
It is known that density functional theory studies\cite{Xiao,Freeman,Stocks} of TIs based on local or semi-local exchange-correlation functionals
can incorrectly predict a material to be a TI when in reality it is not \cite{Zunger}, to the detriment of experiment-theory interplay.
This is usually due to the fact that such a local or semi-local treatment systematically underestimates the band gap
and can lead to an incorrect ordering of the frontier bands at the TRIM points,
which is the determining factor in the prediction of the topological phase.
Furthermore, it can also give rise to wrong band topologies and effective masses\cite{Kim}.
Thus, we have also carried out calculations employing the Modified Becke-Johnson Local Density Approximation (MBJLDA) method\cite{Tran},
which predicts\cite{Kim} band gaps, effective masses and, most importantly, frontier-band ordering
that are in very good agreement with the computationally more intense GW\cite{Svane} and hybrid-functional\cite{Scuseria} approaches.

\begin{figure}[tbp]
\includegraphics[width=1.0\linewidth]{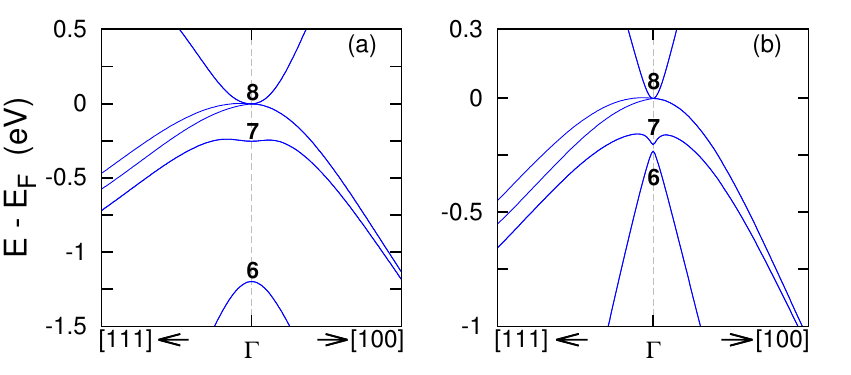}
\caption{Band structure relative to the Fermi energy of unstrained HgSe along the $\Gamma$-X ([100]) and $\Gamma$-L ([111]) symmetry
directions of the BZ [$\vec{k}_L = \frac{\pi}{a} (1, 1, 1)$ and  $\vec{k}_X = \frac{\pi}{a} (1, 0, 0)$, where $a$ is the equilibrium lattice constant]
calculated by (a) the PBE and (b) the MBJLDA approach\cite{Tran}, $c_{\small MBJ} = 1.2$.
}
\label{bandstruct0}
\end{figure}

Figures \ref{bandstruct0}(a) and \ref{bandstruct0}(b) display the PBE and MBJLDA band structures of the un-deformed HgSe, respectively, along two symmetry directions,
where the MBJLDA parameter c=1.2.
PBE predicts the correct level ordering in the upper valence-band region with $E_0$ = -1.2 eV and $\Delta_{SO}$ = 0.25 eV,
thus yielding a $\Gamma_6$ state which is lying far low relative to the $\Gamma_8$ state.
On the other hand, the MBJLDA shifts the $\Gamma_6$ state to a higher energy,
resulting in $E_0$ = -0.23 eV and $\Delta_{SO}$ = 0.20 eV, in agreement with
the values of $E_0$ = -0.32 eV and $\Delta_{SO}$ = 0.27 eV of recent GW calculations\cite{Svane}.
The fact that both the PBE and the LDA\cite{Svane} exchange correlation functionals for HgSe yield the correct frontier band ordering ($\Gamma_7$, $\Gamma_8$)
allows for a reliable determination of the topological invariant\cite{Zhang}. Thus, throughout the remainder of this paper the electronic structure calculations
under strain are carried out using the PBE exchange correlation functional. The atomic positions were fully relaxed
using the conjugate gradient algorithm until all interatomic forces are smaller than 0.01 eV/nm.

\section{Results and Discussion}

\subsection{Bulk band structure under uniaxial strain}
In order to open a band gap in the semi-metallic HgSe, one needs to employ strain.
We have carried out calculations under \textit{hydrostatic} pressure in a wide volume range, from -30\% to +30\% of the equilibrium volume $V_0$,
 in which the equilibrium structure of HgSe remains zinc-blende\cite{Radescu}.
We find that within this wide range HgSe always has a zero band gap and hence remains a semimetal.

Therefore, we have investigated the evolution of the electronic structure and the topological invariant under \textit{uniaxial} strain,
$\epsilon_{[hkl]} \equiv \frac{c-a}{a}$, along the [001] and [110] directions which lowers the $T_d$ symmetry of the ZB structure to $D_{2d}$ and C$_{2v}$, respectively,
and opens a band gap near the $\Gamma$ point.
In both cases, the in-plane lattice constant perpendicular to the strain was fixed to that of the equilibrium ZB structure.
Our ab initio calculations of the total energy as a function of both uniaxial strains indicate that the corresponding structures are mechanically stable up to strain of $\pm$15\%.

\subsubsection{[001] strain of D$_{2d}$  symmetry}
\begin{figure}[tbp]
\includegraphics[width=1.0\linewidth]{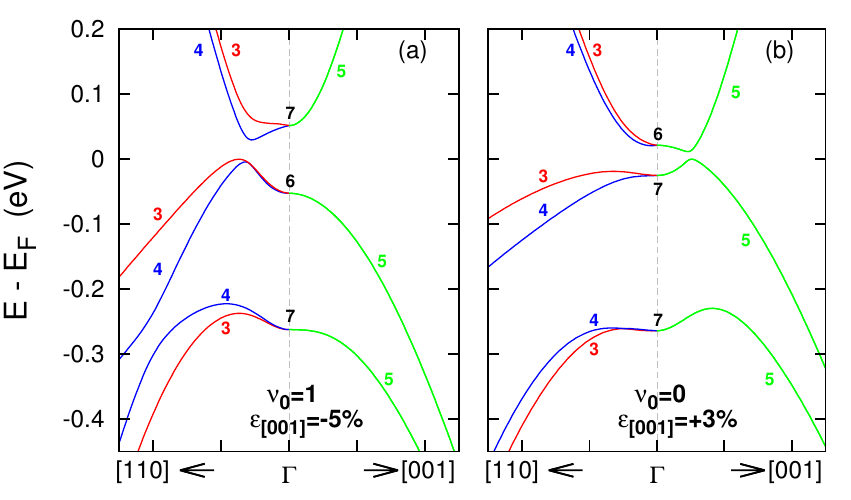}
\caption[ ]{(Color online) Band structure of HgSe (relative to the Fermi energy $E_F$) under uniaxial strain along the [001] direction of D$_{2d}$  symmetry near the $\Gamma$ point along the symmetry directions parallel and
perpendicular to the strain direction.
(a) The compressive strain of $\epsilon_{[001]}$ = -5\% yields a TI ($\nu_0$ = 1) with $E_{gap}$ = 37 meV.
(b) The  tensile strain of $\epsilon_{[001]}$ = +3\% yields a NI ($\nu_0$ = 0) with $E_{gap}$ = 13 meV.
The red and blue color denote the two spin polarized bands, while the green refers to non-spin polarized bands.}
\label{bandstrained001}
\end{figure}

Figs. \ref{bandstrained001}(a) and (b) show the band structures of HgSe under $\epsilon_{[001]}$ of -5\% and +3\%, respectively,
perpendicular to the strain direction.
Under a [001] uniaxial tensile (compressive) strain $\epsilon_{[001]} >(<)0$, of $D_{2d}$ symmetry,
the half-filled four-fold degenerate $\Gamma_{8}$ state splits into an  two-fold degenerate occupied valence $\Gamma_{7v}$ ($\Gamma_{6v}$)
and a  two-fold degenerate unoccupied conduction $\Gamma_{6c}$ ($\Gamma_{7c}$) energy levels leading to a positive (negative) splitting,
$\Delta^\Gamma_{[001]} \equiv E(\Gamma_{6}) - E(\Gamma_{7})>(<)$0.

Under compressive strain ($\Delta^\Gamma_{[001]} <$ 0) the maxima of the two highest non-degenerate valence bands
along the [110] direction, corresponding to the irreducible representations 3 and 4, respectively,
occur away from $\Gamma$. This gives rise to a dumbbell- or camel-back-shape valence bands with
 valence band minima at $\Gamma$,
due to the fact that band energies along [\={1}\={1}0] (not shown in the figure) are equal to those along [110] due to time reversal symmetry.
Similarly, the $\Sigma_4$ conduction band displays a reverse camel-back shape with a maximum at $\Gamma$ and a minimum along the $\Sigma$ direction away from $\Gamma$.
These valence bands along $\Sigma$ become double-degenerate at $\Gamma$.
 This leads to an indirect gap of $E_{gap}$ = 37 meV located in a direction perpendicular to the strain axis, in contrast to the case of the [001] tensile stress
 in Fig. \ref{bandstrained001} (b), where the energy gap $E_{gap}$ occurs along the strain axis.

Fig. \ref{bandstrained001}(b) shows the band structure of HgSe under tensile strain of $\epsilon_{[001]}$ = +3\%  ($\Delta^\Gamma_{[001]} >$ 0)
along the high symmetry [110] and [001] directions parallel and perpendicular to the strain direction, respectively.
Note that along the $\Delta$ ([001]) direction of C$_{2v}$ symmetry
both the conduction and valence bands are  doubly degenerate (denoted by the green curves) of $\Delta_5$ representation.
The crystal-field-splitting gap, $E_{gap}$ = 13 meV, which is allowed by the same symmetry of the bands\cite{Moon}, is observed away from $\Gamma$.

\subsubsection{[110] strain of C$_{2v}$ symmetry}
Application of uniaxial stress along the [110] reduces the symmetry further to the orthorhombic C$_{2v}$.
As this space group does not allow four-fold-generate states,
the ZB $\Gamma_8$ splits under strain and creates a gap between a valence $\Gamma_{5v}$ and conduction $\Gamma_{5c}$,
both of which are two-fold degenerate.
Under expansion the frontier energy bands exchange their ordering.

\begin{figure}[tbp]
\includegraphics[width=1.0\linewidth]{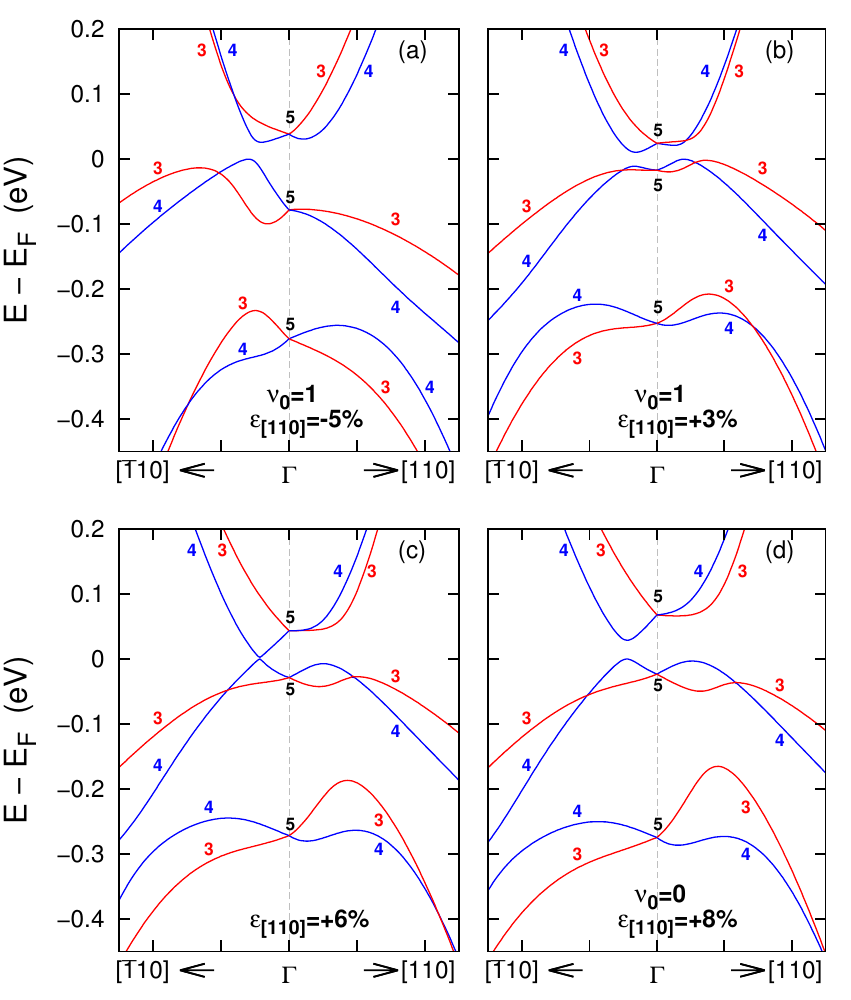}
\caption[ ]{(Color online) Band structure of HgSe under uniaxial strain along the [110] direction of C$_{2v}$ symmetry
near the $\Gamma$ point along the high symmetry directions parallel and perpendicular to the strain direction.
(a) The compressive strain of $\epsilon_{[110]}$ = $-$5\% yields a TI ($\nu_0 = 1$) with $E_{gap}$ = 31 meV.
(b) The tensile strain of $\epsilon_{[110]}$ = +3\% yields a TI ($\nu_0 = 1$) with $E_{gap}$ = 9 meV.
(c) Critical tensile strain $\epsilon_{[110]}$ = +6\% where the conduction and valence bands of irreducible representation 4 cross at E$_F$
and HgSe undergoes a TI $\leftrightarrow$ NI transition.
(d) At larger tensile strain of $\epsilon_{[110]}$ = +8\% HgSe becomes a NI ($\nu_0$ = 0) with a band gap $E_{gap}$ = 25 meV.
The red and blue color denote the spin polarized bands.}
\label{bandstrained110}
\end{figure}

Figs. \ref{bandstrained110}(a)-(d) show the evolution of the band structure around $\Gamma$, under a uniaxial [110] strain of (a) $-$5\%, (b) +3\%, (c) +6\%
and (d) +8\%, along the [110]  ($\hat{x}+\hat{y}$) and $[\bar{1}10]$ ($-\hat{x} + \hat{y}$) high $\vec{k}$ symmetry directions,
which are parallel and perpendicular to the strain direction, respectively.
Under $\epsilon_{[110]}$ = $-$5\% the direct gap at $\Gamma$ is $\Delta^\Gamma_{[110]} \equiv E(\Gamma_{5c}) - E(\Gamma_{5v})$ = 116 meV.
The minimum band gap $E_{gap}$ = 31 meV occurs perpendicular to the strain direction between the camel-back-shape valence and inverse camel-back-shape
conduction bands of $\Sigma_4$ irreducible representation.

\begin{figure}[tbp]
\includegraphics[width=1.0\linewidth]{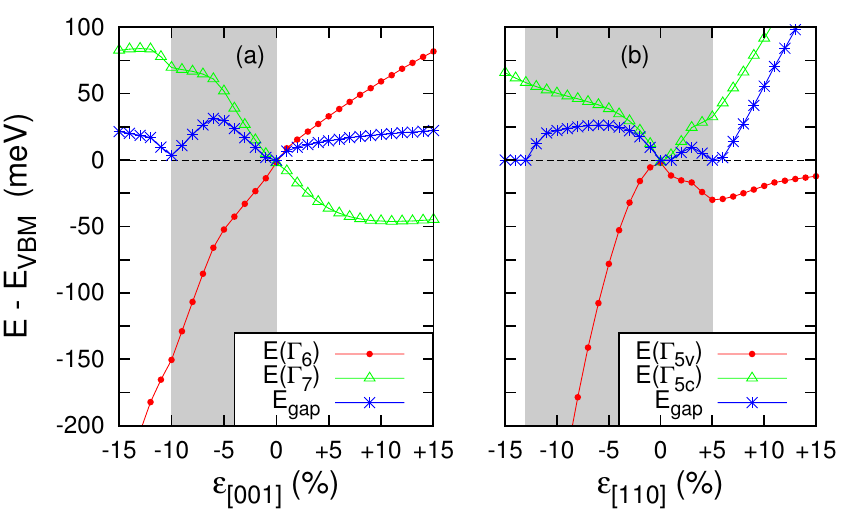}
\caption{(Color online)
Variation of the band gap $E_{gap}$ and the energies of the conduction band and valence band states,
$E(\Gamma_n)$ at the $\Gamma$ point as a function of uniaxial strain along
(a) the [001] direction of D$_{2d}$ symmetry (where $n$= $6$ and $7$) and
(b) the [110] direction of C$_{2v}$ symmetry (where $n$ = $5c$  and $5v$).
The gray shaded areas denote the non-trivial topological phase, where $\nu_0 = 1$.}
\label{Elevels}
\end{figure}

Figures \ref{Elevels}(a) and (b) display the evolution of the band gap $E_{gap}$ and the
energies of the conduction and valence band states $E(\Gamma_n)$ at $\Gamma$ as a function of uniaxial
strain ($-15\% \leq \epsilon \leq +15\%$) applied along the [001] and [110] directions, respectively.
The gray shaded areas denote the strong topological insulating phase with $\nu_0 = 1$,
while the white background corresponds to the normal insulating phase ($\nu_0 = 0$).
The evolution of $\nu_0$ with uniaxial strain is discussed in more detail in the next section.
For [001] strain, the frontier valence and conduction bands at $\Gamma$ are of $\Gamma_6$ and $\Gamma_7$ symmetry,
while they are of $\Gamma_5$ symmetry for the [110] strain.
Under both types of uniaxial compression $E_{gap}$ exhibits a non-monotonic behavior with strain, reaching
 its maximum value in the topological non-trivial phase ($\nu_0 = 1$) at $\epsilon_{[001]}$ = -5\% and $\epsilon_{[110]}$ = -6\%,
and vanishing at some critical strain of $\epsilon_{[001]}$ = -10\% and $\epsilon_{[110]}$ = -13\%, respectively, where the system
undergoes a transition into the trivial insulator phase.  On the other hand, HgSe has a different response between [001] and [110] tensile ($\epsilon >$0)
strain. Namely, for $\epsilon_{[001]} >$ 0 the band gap is trivial ($\nu_0 = 0$) and increases with strain [Fig. \ref{Elevels}(a)] , while
$\epsilon_{[110]} >$ 0 the system is in a strong topological insulator phase ($\nu_0 = 1$) for $\epsilon_{[110]}\leq$ +5\% and
it becomes a trivial insulator ($\nu_0 = 0$) at  larger strain [Fig. \ref{Elevels}(b)].

\subsection{Evolution of Topological Invariant}
An intriguing question is whether there is a correlation between the {\it change} of the frontier valence and conduction bands and corresponding
{\it changes} in the topological invariant $\nu_0$.
Adiabatic deformations of the band structure (not involving band touching or overlaps across the band gap) under external perturbations (such as strain) leave the topology invariant.
Murakami\cite{Murakami} has recently studied the classification of all the possible gap closing in 2D and 3D and showed that
the gap closings is associated with the change of the topological invariant number.

\begin{figure}[tbp]
\includegraphics[width=1.0\linewidth]{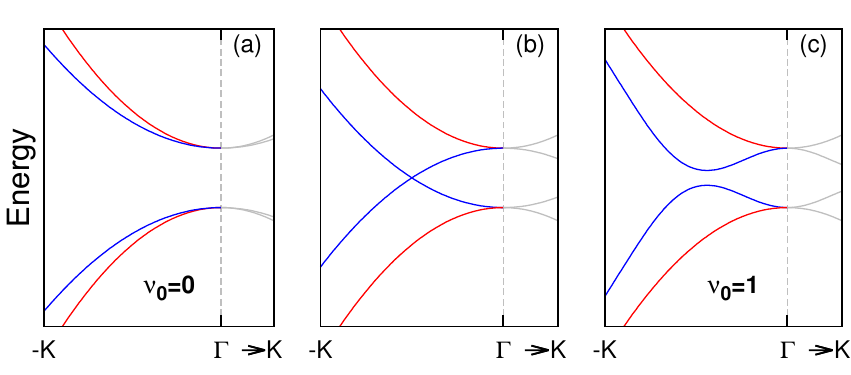}
\caption[ ]{(Color online) Schematic of the strain-induced evolution of the band structure
along an arbitrary direction ($\Gamma \leftrightarrow \pm K$) in the $k_z$=0 plane where the spin-split bands are denoted by red and blue.
(a)The strain splits the originally four-fold degenerate frontier bands at $\Gamma$ in electrons and holes bands which does not
involve any band crossing and hence the bulk system is in a topologically trivial phase with $z_0=0$.
(b) Band inversion of one of the spin-split bands (blue) at the critical strain.
(c) Odd number (one) of band inversions for one of the bands (blue) at a larger strain giving rise to a band gap opening away from $\Gamma$ and the concomitant formation
of a camel-back- (inverse camel-back-) shape valence (conduction) band along the $\Gamma \leftrightarrow \pm K$ direction, rendering the system a strong TI ($z_0=1$).
Note that the inverted bands exhibit both electron and hole character.}
\label{inversion}
\end{figure}

In order to elucidate the underlying mechanism of the evolution of both the band inversion and $\nu_0$, we show schematically in
Fig. \ref{inversion}  the strain-induced change of the band structure close to the TRIM $\Gamma$ point along an
arbitrary direction ($\Gamma \leftrightarrow \pm K$) in the $k_z$=0 plane,
where the spin-split bands are denoted by red and blue. Fig. \ref{inversion}(a) shows that under, for example, a uniaxial tensile strain,
the originally four-fold degenerate $\Gamma_8$ state (under zero strain) splits in the spin-polarized electron and hole
bands without involving any band inversion. This corresponds to the band structure along the [110] direction for $\epsilon_{[001]}>0$
in Fig. \ref{bandstrained001}(b).
In order to reach the topological phase and its characteristic counterpropagation of opposite spin
states (i.e. $\vec{S}_{-\vec{k}} = - \vec{S}_{\vec{k}}$) on the surface, it is necessary to ``knot'' (entangle)
the frontier bands.\cite{KaneMoore,QiZhang} That is achieved by inverting an odd (one) number of spin-split bands (blue) in Fig. \ref{inversion}(b).
Further increase of the strain [Fig. \ref{inversion}(c)] gives rise to the opening
 of a (crystal field)  band gap  with the concomitant formation
of a camel-back- (inverse camel-back-) shape valence (conduction) band along the $\Gamma \leftrightarrow \pm K$ direction.
This corresponds to the non-degenerate $\Sigma_4$ valence and conduction bands along the $[\bar{1}10]$ direction for $\epsilon_{[110]}<0$ in
Fig. \ref{bandstrained110}(a).
Since no other band inversions occur along the [110] direction, the systems is a strong topological insulator ($z_0$=1).
Similarly,  there is only one band inversion of the valence and conduction bands of $\Sigma_4$ character
along the $[\bar{1}10]$ direction for $\epsilon_{[110]}>$0 in Fig. \ref{bandstruct0}(b), rendering the system a strong TI.
Moreover, the doubly-degenerate valence and conduction bands of $\Delta_5$  character along the [001] direction for
$\epsilon_{[001]} > 0$  [Fig. \ref{bandstrained110}(b)] have an even (two) number of band
inversions and hence the system is a topologically a trivial insulator (z$_0$=0).

While the above mechanism, based on the changes of the frontier electronic bands (eigenvalues), provides an intuitive
explanation of the topological phase transitions, it requires knowledge of the
overlap of the 2D conduction and valence energy surfaces. It becomes less practical when these surfaces are not smooth and present multiple extrema.
That is the case for Figs. \ref{bandstrained110}(b)-(d), for instance, requiring a careful examination along multiple $\vec{k}$-directions.
For such cases, a less intuitive, but more stringent criterion, is the determination of the topological invariant, $\nu_0 \equiv (z_0 - z_\pi)$ mod 2,
based on the evolution of the Wannier functions (which include a phase factor), as explained in section II.
Nonetheless, one notes a clear trend in the evolution of the bands along the $[\bar{1}10]$ direction.
The gap between the valence and conduction bands of irreducible representation 4 is progressively reduced with increasing tensile strain
until they cross at $\sim$0 eV for $\epsilon_{[110]}= 5-6\%$ [Fig. \ref{bandstrained110}(c)]. The system undergoes a
second band inversion that ``unknots'' the band structure.
As expected, this closing of the gap coincides with the topological phase transition predicted by the invariant $\nu_0$.
For larger strain, this gap increases monotonically [Fig. \ref{bandstrained110}(d)] and without changes to the topological phase.

In HgSe, band inversions occur only in the vicinity of $\Gamma$, because the conduction and valence states at other TRIM points are too far separated in energy.
Consequently, the bands in the $k_z$=$\pi$ plane are always not inverted [analogous to Fig. \ref{inversion}(a)] and hence
$z_\pi$=0 in the entire strain range, thus yielding $\nu_0 = (z_\pi - z_0 \text{ mod } 2) = z_0$.

\begin{figure}[tbp]
\includegraphics[width=1.0\linewidth]{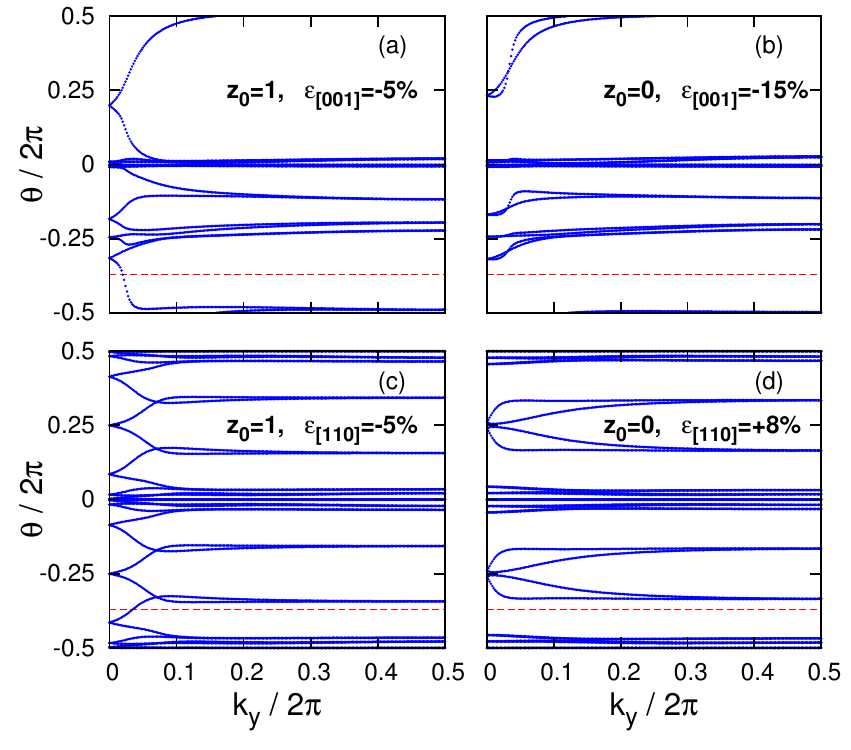}
\caption{(Color online)
Evolution of Wannier centers for HgSe along $k_y$ in the $k_z$=0 plane under uniaxial strain along the [001] direction of
(a) $\epsilon_{[001]}$ = -5\%  and (b) $\epsilon_{[001]}$ = +15\%, and along the [110] direction of
(c) $\epsilon_{[110]}$ = -5\%  and (b) $\epsilon_{[110]}$ = +8\%, respectively.
The evolution lines cross any arbitrary reference line parallel to $k_y$ (for example the red dotted line) an odd (even) number of times yielding $z_0$ = 1 (0).
In all cases the corresponding $z_{\pi}$ = 0.
$N_x = 48$.}
\label{Z2}
\end{figure}

Figures \ref{Z2}(a) and \ref{Z2}(b) show the evolution lines of the Wannier centers along $k_y$ in the $k_z$=0 plane under a uniaxial compression
of $\epsilon_{[001]}$ = -5\% and $\epsilon_{[001]}$ = -15\% along the [001] direction, respectively.
Under the -5\% (-15\%) compression the evolution curves (blue) cross any arbitrary line parallel to the horizontal axis
(for example the red dotted line) an odd (even) number of points, thus yielding $z_0$ = 1 ($z_0$ = 0).
Furthermore, the evolution lines of the Wannier centers along $k_y$ in the
$k_z$=$\pi$ plane under both uniaxial compression (not shown here) yield $z_{\pi}$ = 0.
This demonstrates that under uniaxial compression along [001],
there is a topological difference between the the TI phase ($\nu_0$=1) at $\epsilon_{[001]}$ = $-$5\%
and the trivial insulating phase ($\nu_0$=0) at $\epsilon_{[001]}$ = $-$15\%.
Interestingly, the critical value of $\epsilon_{[001]}$ = $-$10\% where HgSe undergoes a TI $\leftrightarrow$ NI transition is close to
the critical value of $\epsilon\sim-11\%$ where the ZB structure undergoes a transition to the cinnabar structure\cite{Radescu}.
In Figures \ref{Z2}(c) and \ref{Z2}(d) we show the evolution of the Wannier centers in the $k_z$=0 plane under a uniaxial compression
($\epsilon_{[110]}$ = $-$5\%) and expansion ($\epsilon_{[110]}$ = +8\%) along the [110] direction, where $z_0$ = 1  and $z_0$ = 0, respectively.
Since in both cases $z_{\pi}$ = 0, HgSe is a TI ($\nu_0$ = 1) under $-$5\% uniaxial compression while it is in the trivial insulating phase ($\nu_0$ = 0)
under 8\% uniaxial expansion.
\subsection{Surface band structure and spin polarization}
\begin{figure}[tbp]
\includegraphics[width=1.0\linewidth]{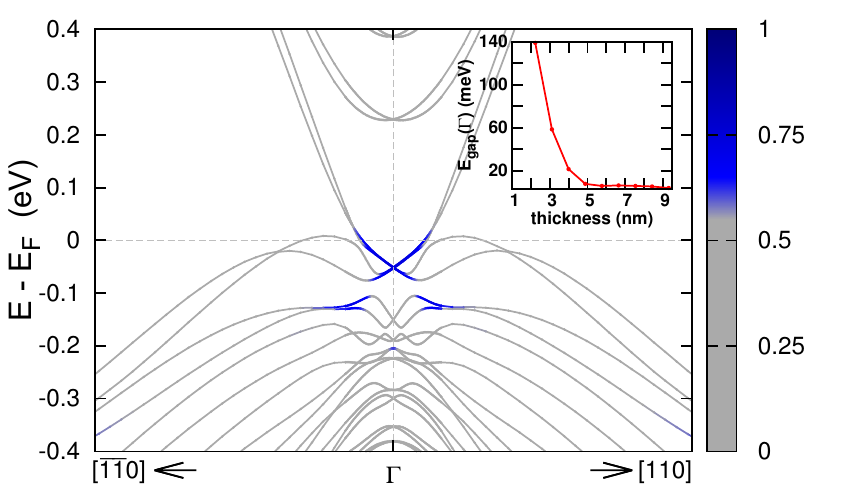}
\caption{(Color online)
Band structure of the (110) HgSe surface under a [110] uniaxial compressive strain ($\epsilon_{[110]}$ = $-$10\%)
along the symmetry directions in the two-dimensional BZ showing a single Dirac-cone at $\Gamma$ at 52 meV below the Fermi energy.
The gray and blue colored bands correspond to bulk- and  surface-derived states,
where the latter are associated with the two top and bottom atomic layers\cite{Park}.
}
\label{slab}
\end{figure}
While the $\mathbb{Z}_2$ criterion is sufficient to ascertain whether a 3D bulk system is a TI,
an additional commonly employed criterion is the existence of gapless surface states that have spin texture.
Therefore, we have carried out electronic structure calculations of the non-polar (110) HgSe surface,
to minimize the effects of surface charges and reconstruction associated with polar surfaces.
The band structure of a  uniaxially strained ($\epsilon_{[110]}$ = -10\%) HgSe (110) slab is shown in Fig. \ref{slab} along the symmetry directions in the 2D BZ.
The states marked in blue are identified to be spatially confined to the top and bottom boundary layers of the slab,
i.e. surface states, using the approach of Park {\it et al}\cite{Park}.
The 21 atomic layer slab exhibits a mirror symmetry with respect to both surfaces.
One can clearly see that the topological surface states form two superimposed Dirac cones at the $\Gamma$ point,
each spatially located on the top and bottom surfaces.
The Dirac point (52 meV below the Fermi energy) is four-fold degenerate due to Kramer's degeneracy at the TRIM point $\Gamma$.
The inset shows the variation of the energy gap of the slab at $\Gamma$ as function of the slab thickness, $D = Na/\sqrt{2}$,
where $N$ is the number of atomic layers and $a$ = 6.278 \AA~is the equilibrium lattice constant. The large
value of the energy gap in thin slabs is due to the interactions between the states localized at the opposite surfaces.
As the number of layers increases, the size of the gap decreases converging to $\sim$ 1 meV at the critical thickness of about 7 nm
($N \sim$ 16).

\begin{figure}[tbp]
\includegraphics[width=1.0\linewidth]{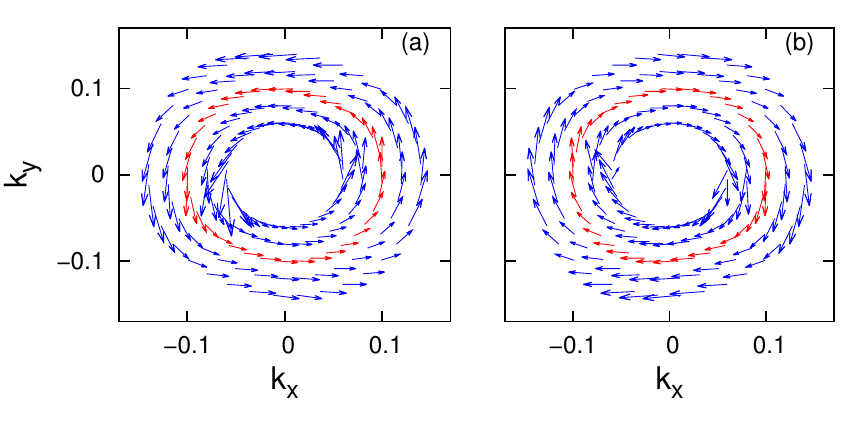}
\caption{(Color online)
Constant energy contours of surface spin polarization vector of the uniaxially strained (110) HgSe surface in Fig. \ref{slab}
within an energy of $\pm$25 meV about $E_F$ (all state above the Dirac point), projected onto the (a) top and (b) bottom (011) HgSe surfaces, respectively.
The red arrow contours denote the spin polarization vector of states at $E_F$ separating those with energies higher and lower than $E_F$.
The units of $k$ are $2\pi/a$.}
\label{spin}
\end{figure}

Fig. \ref{spin} shows constant energy contours of the surface spin polarization vector projected on the $k_x-k_y$ plane
in the vicinity of the Fermi energy on the top (a) and bottom (b) surface layers, respectively.
The red arrow contours denote the spin polarization vector of states at $E_F$,
with the remaining energy contours representing an energy range of $\pm$25 meV about the Fermi energy.
We find that $\vec{S}_{-\vec{k}_{||}} \approx - \vec{S}_{\vec{k}_{||}}$ and $\vec{S}_{\vec{k}_{||}}$ is locked almost normal to the in-plane momentum $\vec{k}_{||}$,
which results from a combination of the strong SOC and the inversion symmetry breaking at the surface, the well-known Rashba effect.\cite{Rashba} This in turn results
 in the rotation of the spin orientation around the Fermi surface (spin chirality) as reported for other known TI materials.

\section{Conclusions}
In conclusion, we have investigated the evolution of the band structure and of the topological invariant in HgSe under application
of [001] and [110] uniaxial strain.
We predict that HgSe is a strong TI ($\nu_0$ =1) in the compressive strain range, $\epsilon^{cr}_{[001]([110])}\leq\epsilon_{[001]([110])} \leq0$, where
the (crystal field) band gap occurring away from $\Gamma$, displays a non-monotonic strain behavior. The band gap vanishes at
 $\epsilon^{cr}_{[001]([110])}$, where the system undergoes a TI $\leftrightarrow$ NI phase transition.
We find that HgSe exhibits a similar behavior applying a [110] tensile strain. On the other hand, under [001] tensile strain HgSe remains a NI.
Thus, these calculations demonstrate  that HgSe can be tuned into a 3D topological insulator via proper strain engineering.

\section{Acknowledgments}
The research at CSUN was supported by NSF-PREM Grant No. DMR-1205734.
Jin Li was supported by a DTRA Grant No. HDTRA1-10-1-0113. YPC acknowledges support by the DARPA MESO program.

\end{document}